\documentclass[preprint,preprintnumbers,aps,superscriptaddress,nofootinbib,tightenlines,floatfix]{revtex4}

\usepackage{amsmath,amssymb}
\usepackage{graphicx}
\usepackage{bm}
\usepackage{comment}

\newcommand{\beq}{\begin{equation}}
\newcommand{\eeq}{\end{equation}}
\newcommand{\bea}{\begin{eqnarray}}
\newcommand{\eea}{\end{eqnarray}}

\def\Nprops{292,500} 
\def\Ncfgs{1194}

\def\mpi{m_\pi}
\def\mK{m_K}

\def\OMIT#1{{}}

\newcommand{\lsim}{\raisebox{-0.7ex}{$\stackrel{\textstyle <}{\sim}$ }}

\newcount\hour \newcount\hourminute \newcount\minute 
\hour=\time \divide \hour by 60
\hourminute=\hour \multiply \hourminute by 60
\minute=\time \advance \minute by -\hourminute
\newcommand{\mydate}{\ \today \ - \number\hour :\number\minute}

\begin{document}
\begin{figure}[!t]

  \vskip -1.5cm
  \leftline{\includegraphics[width=0.25\textwidth]{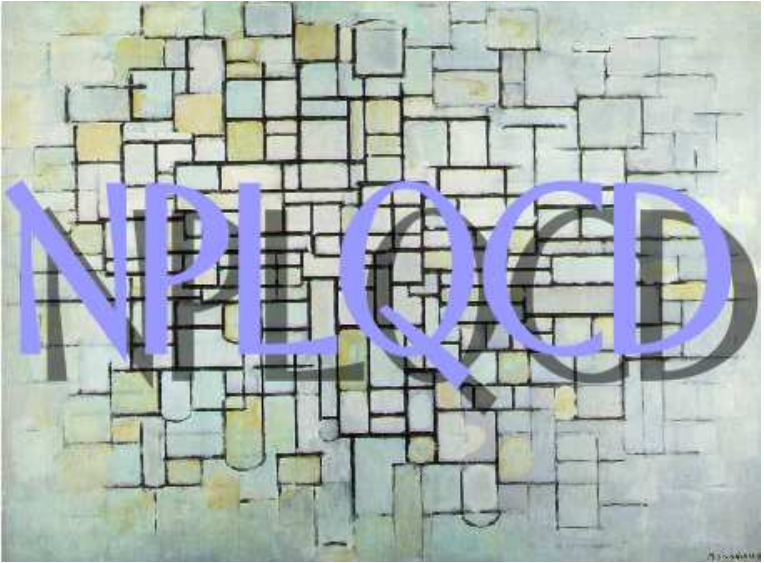}}
\end{figure}

\preprint{\vbox{ 
\hbox{UNH-09-02} 
\hbox{JLAB-THY-09-984}
\hbox{NT@UW-09-09} 
\hbox{ATHENA-PUB-09-016}
}}

\vskip .5cm

\title{High Statistics Analysis using Anisotropic Clover Lattices: (II)
   Three-Baryon Systems}

\vskip .5cm \author{Silas R.~Beane} \affiliation{Department of
  Physics, University of New Hampshire, Durham, NH 03824-3568.}
\author{William Detmold} \affiliation{Department of Physics, College
  of William and Mary, Williamsburg, VA 23187-8795.}
\affiliation{Jefferson Laboratory, 12000 Jefferson Avenue, Newport
  News, VA 23606.}  \author{Thomas C.~Luu} \affiliation{N Division,
  Lawrence Livermore National Laboratory, Livermore, CA 94551.}
\author{Kostas Orginos} \affiliation{Department of Physics, College of
  William and Mary, Williamsburg, VA 23187-8795.}
\affiliation{Jefferson Laboratory, 12000 Jefferson Avenue, Newport
  News, VA 23606.}  
\author{Assumpta Parre\~no}
\affiliation{Departament d'Estructura i Constituents de la Mat\`{e}ria
  and Institut de Ci\`encies del Cosmos, Universitat de Barcelona,
  E--08028 Barcelona, Spain.}  
\author{Martin J.~Savage}
\affiliation{Department of Physics, University of Washington, Seattle,
  WA 98195-1560.}  \author{Aaron Torok} \affiliation{Department of
  Physics, University of New Hampshire, Durham, NH 03824-3568.}
\author{Andr\'e Walker-Loud} \affiliation{Department of Physics,
  College of William and Mary, Williamsburg, VA 23187-8795.}
\collaboration{ NPLQCD Collaboration } \noaffiliation \vphantom{}

\date{\mydate}

\vskip 0.8cm
\begin{abstract}
  \noindent
  We present the results of an exploratory Lattice QCD calculation of
  three-baryon systems through a high-statistics study of one ensemble
  of anisotropic clover gauge-field configurations with a pion mass of
  $m_\pi\sim 390~{\rm MeV}$.  Because of the computational cost of the
  necessary contractions, we focus on correlation functions generated
  by interpolating-operators with the quantum numbers of the
  $\Xi^0\Xi^0 n$ system, one of the least demanding three baryon
  systems in terms of the number of contractions.  We find that the
  ground state of this system has an energy of $E_{\Xi^0\Xi^0n}=
  3877.9\pm 6.9\pm 9.2\pm3.3$~MeV corresponding to an energy-shift due
  to interactions of $\delta
  E_{\Xi^0\Xi^0n}=E_{\Xi^0\Xi^0n}-2M_{\Xi^0} -M_n=4.6\pm 5.0\pm
  7.9\pm4.2~{\rm MeV}$.  There are a significant number of time-slices
  in the three-baryon correlation function for which the
  signal-to-noise ratio is only slowly degrading with time.  This is
  in contrast to the exponential degradation of the signal-to-noise
  ratio that is observed at larger times, and is due to the suppressed
  overlap of the source and sink interpolating-operators that are
  associated with the variance of the three-baryon correlation
  function onto the lightest eigenstates in the lattice volume
  (mesonic systems).  As one of the motivations for this area of
  exploration is the calculation of the structure and reactions of
  light nuclei, we also present initial results for a system with the
  quantum numbers of the triton ($pnn$).  This present work
  establishes a path to multi-baryon systems, and shows that Lattice
  QCD calculations of the properties and interactions of systems
  containing four and five baryons are now within sight.
\end{abstract}
\maketitle




%
%
\section{Introduction \label{sec:Intro} }
\noindent
One of the ultimate goals of Lattice Quantum Chromodynamics (LQCD) is to calculate the
properties and interactions of light-nuclei to high precision from
first principles.  While it is important to be able to explicitly
demonstrate that nuclei emerge from QCD, the underlying motivation for
this effort is to provide proof that LQCD provides a reliable theoretical
tool with which to calculate highly complex low-energy strong
interaction processes.  With such a tool in hand, calculations of
strong interaction systems for which experimental guidance is minimal,
or absent, can be performed with confidence and with uncertainties
that can be rigorously quantified.  
The interaction between three neutrons, which is an important input
into many-body calculations of nuclei, provides an example of a quantity that
is difficult to access experimentally, but which will be calculable to high
precision with LQCD within the next decade.

The theoretical framework with which to determine the hadron-hadron
scattering phase-shifts below the inelastic threshold from the volume
dependence of the two-hadron energy levels in the lattice-volume was
established a number of years ago by
L\"uscher~\cite{Luscher:1986pf,Luscher:1990ux}.  This framework was
used to extract nucleon-nucleon scattering lengths in quenched QCD
(QQCD)~\cite{Fukugita:1994ve} at unphysically large pion masses.
Subsequent fully-dynamical LQCD calculations also used the
L\"uscher-method to extract nucleon-nucleon~\cite{Beane:2006mx}, and
hyperon-nucleon~\cite{Beane:2006gf} scattering lengths and
phase-shifts~\footnote{Calculations of these same processes were
  subsequently performed in quenched
  QCD~\cite{Ishii:2006ec,Nemura:2008sp} and in QCD
  \cite{Ishii:2009zr,Nemura:2009kc}.  In these same works it was
  suggested that a phenomenologically useful baryon-baryon potential
  could be defined from LQCD calculations.  However, flawed reasoning
  led to such a conclusion, in particular, the omission of the
  spatially-dependent two-body overlap factor which is present in the
  correlation
  functions~\cite{Detmold:2007wk,Beane:2008dv,Beane:2008ia}.}  (from a
single correlation function), albeit at unphysically large pion
masses.  The exponentially degrading signal-to-noise ratio encountered
in the region of the correlation functions dominated by the
ground-state, expected from the arguments presented by
Lepage~\cite{Lepage:1989hd}, severely limited the precision with which
the scattering phase-shifts could be extracted from all of these
calculations.  The resources required to perform these calculations
and the anticipated scaling of the statistical uncertainties in such
calculations as a function of the pion mass~\cite{Lepage:1989hd} are
sufficient to estimate the resources required to perform calculations
of baryon-baryon (BB) scattering at the physical value of the pion
mass to a given level of precision~\cite{Beane:2008dv}.  Given the
smallness of nuclear physics energy-scales, typically a few MeV,
compared to the chiral-symmetry breaking scale, $\Lambda_\chi$, or the
pion mass, $m_\pi$, it is guaranteed that a very large number of
measurements will be required to achieve the necessary precision.  At
the physical pion mass, Ref.~\cite{Beane:2008dv} estimates that $\sim
3\times 10^6$ measurements will be required to determine the NN
scattering length with a $\sim 50\%$ uncertainty.  Such estimates will
be further refined as additional calculations at different pion
masses, lattice-volumes and lattice-spacing are performed.

Recently, we have performed a high-statistics calculation of a number
of single-hadron correlation functions~\cite{Beane:2009ky} on an
ensemble of the anisotropic gauge-field configurations generated by
the Hadron Spectrum Collaboration~\cite{Lin:2008pr,Edwards:2008ja}
with a pion mass of $m_\pi\sim 390~{\rm MeV}$, a spatial lattice
spacing of $b_s\sim 0.1227$, an anisotropy $\xi=b_s/b_t=3.5$ and a lattice
volume of $20^3\times128$.  The goal of the study was to ``jump'' an
order of magnitude in the number of measurements performed to estimate
correlation functions, and to explore the ``new territory'' that
subsequently emerged.  The baryon masses were extracted with
precision at the $\lsim 0.2\%$-level from the $\Nprops$ measurements
performed on $\Ncfgs$ of these gauge-field configurations.  A number
of important and surprising observations were made in that work that
have modified the path from LQCD to nuclei that we envisage.
One of the most important aspects of that high-statistics
work was that a detailed study of the signal-to-noise
ratio in the single-baryon correlation functions became  possible.  
The signal-to-noise ratio was found to be approximately
independent of time for a significant number of time-slices prior to
evolving toward the expected exponential
degradation~\cite{Lepage:1989hd,Beane:2009ky}.  This window of
``clean'' time-slices is understood in terms of the the relative
magnitude of the overlap of the single-baryon interpolating operator
onto the single-baryon eigenstates, compared with the overlap of the
corresponding interpolating-operator onto the lightest eigenstates
(involving both meson and baryon-anti-baryon states) that contribute
to the correlation function that governs the variance of the
single-baryon correlation function.  Given that the signal-to-noise ratio for a
system containing more than one baryon is expected to scale
(approximately) as the product of the signal-to-noise ratio's of the
individual baryons (neglecting their interactions), this window of
clean time-slices suggests that it may well be possible to calculate
the energy-levels of systems containing a number of baryons in this
lattice volume with these interpolating-operators.

In this work we present the first LQCD calculations of system
comprised of three baryons.  As the number of contractions required to
form the correlation functions is naively $N_u! \ N_d!\ N_s!$, one of the
least computationally expensive systems\footnote{The
  $\Lambda\Lambda\Sigma^0$ system requires fewer contractions, however
  the additional strange quark in the $\Xi^0\Xi^0n$ system is expected to result
  in a cleaner signal.} to explore is the one that couples to a source
and sink of the form $\Xi^0\Xi^0 n$. 
For simplicity, the product of the single baryon interpolating operators with the
quantum numbers of the $\Xi^0$ and $n$ are used in the calculations.  This source will
have non-zero overlap with  states with strangeness
four ($s=4$), spin one-half ($J={1\over 2}$), and with a third
component of isospin of one-half ($I_z={1\over 2}$).  Further, we
present preliminary calculations of the $pnn$-system which will contain
the triton if it is bound for this pion mass.  We are presently unable
to explore even the simplest system containing four baryons
because of the computational resources required to perform the
contractions (which are usually the least expensive component of a
lattice calculation!), but see no reason why systems containing four
and five baryons could not be explored in the near future.

\section{Lattice QCD Calculations}
\label{sec:lattice-calculations}

\noindent
In this study, we employ a single ensemble of the $n_f=2+1$-flavor
anisotropic clover gauge-field configurations that have been produced
by the Hadron Spectrum Collaboration~\cite{Lin:2008pr,Edwards:2008ja}.
The technical details of the propagators computed on this ensemble
are presented in Ref.~\cite{Beane:2009ky} and we do not repeat them
here. In the current calculation, 
the analysis is restricted to a
slightly smaller data set, corresponding to an average on 218 randomly
distributed measurements on each of 1191 configurations (a total of
$\sim$260,000 measurements).


Each of the propagators calculated on the gauge-field configurations 
is used to determine  two-point correlation functions,
which for a single baryon have the form
\begin{equation}
  C_{{\cal H}; \Gamma }({\bf p};t)= \sum_{{\bf x}}e^{i{\bf p \cdot x}}\ 
  \Gamma^\alpha_\beta\ 
  \langle\  {\cal H}^\beta({\bf x},t) \overline{\cal H}_\alpha({\bf
    x_0},0)\  \rangle
  \ \ \ ,
  \label{eq:2}
\end{equation}
where ${\cal H}^\alpha({\bf x},t)$ is an interpolating operator for
the appropriate baryon state, {\it e.g.}, for the proton ${\cal
  H}^\alpha({\bf x},t) = \epsilon_{abc} \left(u^{a,T}\ C\ \gamma_5
  d^b\right) u^{c,\alpha}({\bf x},t)$ where $C$ is the charge
conjugation matrix. The Dirac matrix $\Gamma$ is an arbitrary
particle-spin-projector and the point ${\bf x_0}$ is the propagator
source point.  The interpolating-operator at the source,
$\overline{\cal H}$, is constructed from gauge-invariantly-smeared
quark field operators, while at the sink, the interpolating operator
is constructed from either local quark field operators, or from the
same smeared quark field operators used at the source, leading to two
sets of correlation functions.  For brevity, we refer to the two sets
of correlation functions that result from these source and sink
operators as {\it smeared-point} (SP) and {\it smeared-smeared} (SS)
correlation functions, respectively.  The correlation functions for
the three-baryon systems have the form~\footnote{ A more complete
  calculation would generate correlation functions between sources and
  sinks that carry the same global quantum numbers, such as
  $\Xi^0\Xi^0 n\rightarrow \Xi^0 \Sigma^+\Sigma^-$, in order to
  identify all of the states in the lattice volume. For computational
  expediency we study only one combination of source and sink. },
\begin{eqnarray}
 &&  C_{{\cal H}_1{\cal H}_2{\cal H}_3; \tilde\Gamma }({\bf p}_1,{\bf p}_2,{\bf p}_3;t)
\  = \ 
\sum_{{\bf x}_1,{\bf x}_2,{\bf x}_3 }\ 
e^{i{\bf p_1 \cdot x_1}}\ e^{i{\bf p_2 \cdot x_2}}\ e^{i{\bf p_3 \cdot x_3}}\ 
  \tilde\Gamma^{\alpha_1\alpha_2\alpha_3}_{\beta_1\beta_2\beta_3} \ 
\nonumber\\
&&
\qquad
  \langle\  
{\cal H}_1^{\beta_1}({\bf x}_1,t) {\cal H}_2^{\beta_2}({\bf x}_2,t) {\cal H}_3^{\beta_3}({\bf x}_3,t) 
\overline{\cal H}_{1,\alpha_1}({\bf x_0},0)
\overline{\cal H}_{2,\alpha_2}({\bf x_0},0)\overline{\cal H}_{3,\alpha_3}({\bf x_0},0)
\  \rangle
  \ \ \ ,
  \label{eq:BBBcorr}
\end{eqnarray}
where $\tilde\Gamma$ is the tensor that projects onto the required
angular momentum state.  The same quark-propagators have been used in
each baryon, and thus the source for each baryon is located at the
same spatial point.  More physically motivated sources and sinks involving
spatial separations would likely improve the overlap onto the ground
state in these systems, however this approach would be more
computationally demanding and is not used in this exploratory work.

In the present work, we have restricted ourselves to the calculation
of correlation functions for which each baryon is projected to
zero-momentum at the sink, defining $C_{{\cal H}_1{\cal H}_2{\cal
    H}_3; \tilde\Gamma }(t) = C_{{\cal H}_1{\cal H}_2{\cal H}_3;
  \tilde\Gamma }({\bf 0},{\bf 0},{\bf 0};t)$.  Further, the optimal
analysis of the three-baryon systems (with propagators from a single
source) would have involved calculating the correlation functions
associated with the different sink-smearing, SP and SS, for each
baryon.  Due to lack of computational resources we have restricted
ourselves to the (SS)$^3$ and the (SP)$^3$ correlation functions, and
have not calculated the ``mixed'' correlation functions, such as the
(SS)$^2$(SP) correlation function.

With two correlation functions associated with each set of quantum
numbers in both the one- and three-baryon sectors, a linear
combination of the pair of correlation functions can be constructed to
produce a combined correlation function that more cleanly projects
onto the lowest energy state in the lattice volume.  One way to
accomplish this is by hand, where one simply ``looks'' for the linear
combination of correlation functions that has an effective mass plot (EMP) with
the ground-state extending to the shortest time-slice.  A refinement
of this ``brute-force'' method is to use the matrix-Prony method
presented in our previous paper~\cite{Beane:2009ky}, which we now review.

The two correlation functions, SS and SP, are sums of exponentials and 
satisfy the following matrix relation,
\begin{equation}
  M y(\tau+t_J) - V y(\tau) = 0
  \ \ \ ,
  \label{eq:recursion}
\end{equation}
where $M$ and $V$ are $2\times 2$ matrices and $y(t)$ is a column
vector with two components corresponding to the two correlation
functions.  Eq.~(\ref{eq:recursion}) implies then the correlation
functions are
\begin{equation}
  y(t) = \sum_{n=1}^2  A_n q_n \lambda_n^{-t}
  \ \ \ ,
  \label{eq:signal}
\end{equation}
where $q_n$ and $\lambda_n=\exp(m_n)$ are the eigenvectors and
eigenvalues of the following generalized eigenvalue problem
\begin{equation}
  M q = \lambda^{t_J} V q\ \ \ .
  \label{eq:gev}
\end{equation}
Given the two sets of correlation functions, the masses can be found
by determining the matrices $M$ and $V$ that are needed in order for
the signal to satisfy Eq.~(\ref{eq:recursion}). Solving
Eq.~(\ref{eq:gev}), leads to the eigenvalues and eigenvectors
needed to reconstruct the amplitudes of  each exponential
in the correlation functions.  It is straightforward to show that
a solution for $M$ and $V$ is
\begin{eqnarray}
  M &=& \left[\
    \sum_{\tau=t}^{t+t_W} y(\tau+t_J)y(\tau)^T\
  \right]^{-1}
  \ \ \ \ ,\
  \ \ \
  V \ =\  \left[\ \sum_{\tau=t}^{t+t_W} y(\tau)y(\tau)^T \ \right]^{-1}
  \ \ \ .
  \label{eq:solution}
\end{eqnarray}
These inverses exist provided that the range, $t_W$, is large
enough so that the matrices in the brackets are of full rank. Once the
eigenvalues, $\lambda_n$ and eigenvectors $q_n$ are determined, the
amplitudes, $A_n$, can be reconstructed using a fixed time-slice as a
normalization point.  
The parameters $t_W$ and $t_J$ can be used to improve
stability as investigated in Ref.~\cite{Beane:2009ky}.  
The
eigenvectors associated with the ground-state energy eigenvalue
provide the linear combination of SS and SP correlation functions for
which the plateau in the EMP sets in at the earliest time.  As the
eigenvectors can be determined at early time-slices, the degradation
of the signal at later times seen in the eigenvalues of the
matrix-Prony method, largely due to increasing fluctuations in the SS
correlation function, is greatly reduced.  This method is
independently applied to the single baryon and the three-baryon pairs
of correlation functions, to produce a single correlation function for
each.  For the present calculations, the
relevant ``diagonalized'' correlation functions are
\begin{eqnarray}
\overline{C}_{\Xi^0}(t) & = & \eta^{(SS)}_{\Xi^0}\ C^{(SS)}_{\Xi^0}(t)
\ +\  \eta^{(SP)}_{\Xi^0}\ C^{(SP)}_{\Xi^0}(t)
\nonumber\\
\overline{C}_{n}(t) & = & \eta^{(SS)}_{n}\ C^{(SS)}_{n}(t)
\ +\  \eta^{(SP)}_{n}\ C^{(SP)}_{n}(t)
\nonumber\\
\overline{C}_{\Xi^0\Xi^0 n}(t) & = & \eta^{(SS)}_{\Xi^0\Xi^0 n}\ C^{(SS)}_{\Xi^0\Xi^0 n}(t)
\ +\  \eta^{(SP)}_{\Xi^0\Xi^0 n}\ C^{(SP)}_{\Xi^0\Xi^0 n}(t)
 \ \ \ .
  \label{eq:evecs}
\end{eqnarray}
where the coefficients $\eta^{(W)}_{\cal H}$ ($W=SS, SP$)
are determined
numerically.  To present the results, and to extract the energies of
the states, it is convenient to work with the effective mass (EM),
$M(t;t_J)$, defined via the ratio
\begin{eqnarray}
M(t;t_J) & = & {1\over t_J} \log\left[\ {\overline{C}_{\cal H}(t)
\over \overline{C}_{\cal H}(t+t_J)} \ \right]
 \ \ \ ,
  \label{eq:emp}
\end{eqnarray}
which is independent of time when the diagonalized correlation function is
dominated by a single exponential.
\begin{figure}[!t]
  \centering
  \includegraphics[width=0.45\columnwidth]{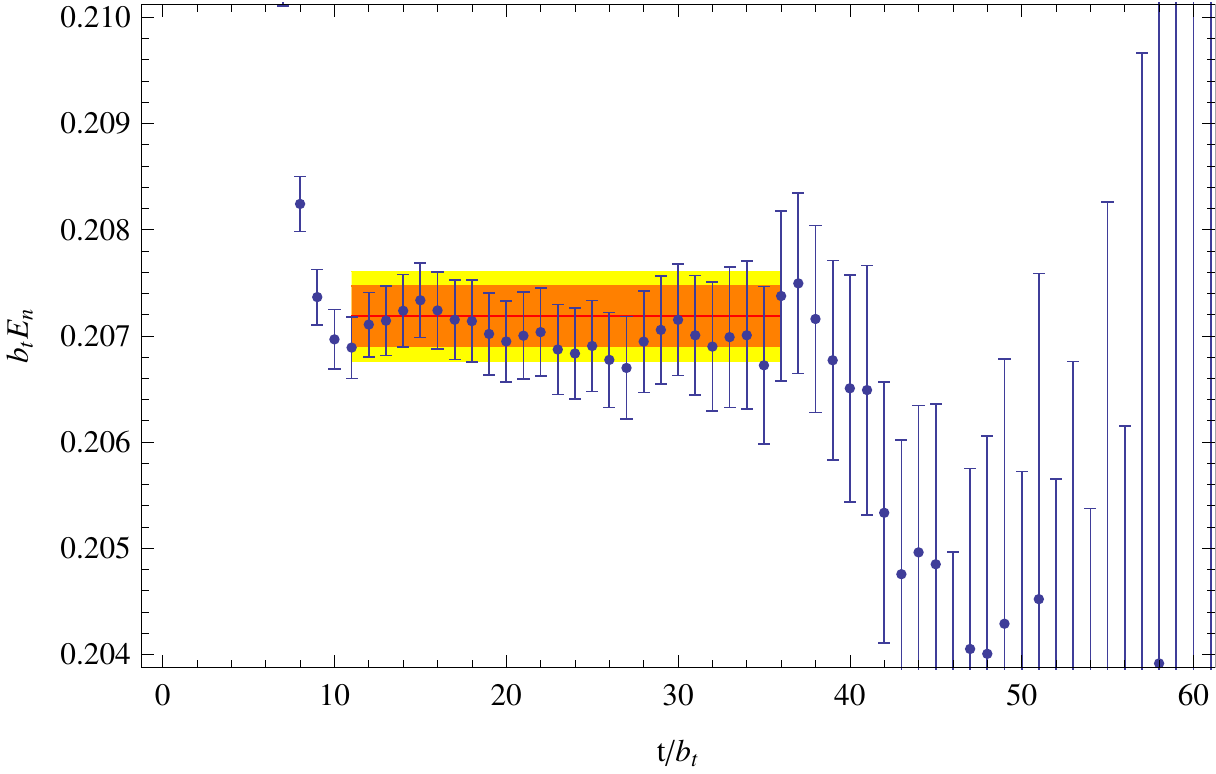}\ \ \ \ \ 
  \includegraphics[width=0.45\columnwidth]{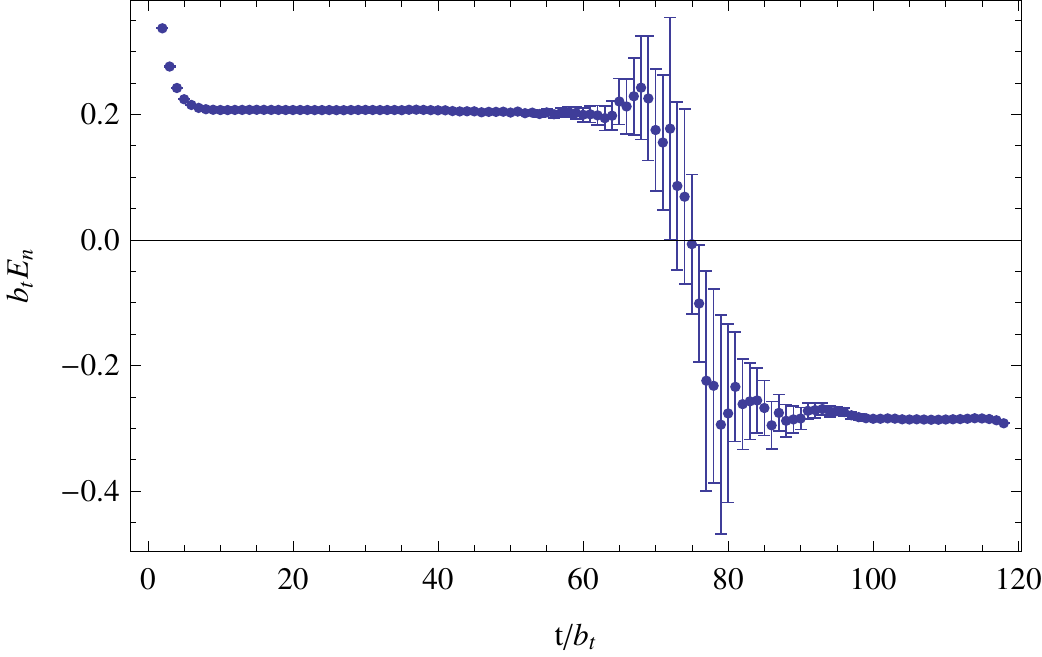}
  \caption{The EMP's associated with the diagonalized nucleon correlation
    function, with $t_J=5$. The left panel, which also shows the fit to the plateau region,  
is a magnification of the right panel.
}
\label{fig:Nemps}
\end{figure}
\begin{figure}[!ht]
  \centering
  \includegraphics[width=0.45\columnwidth]{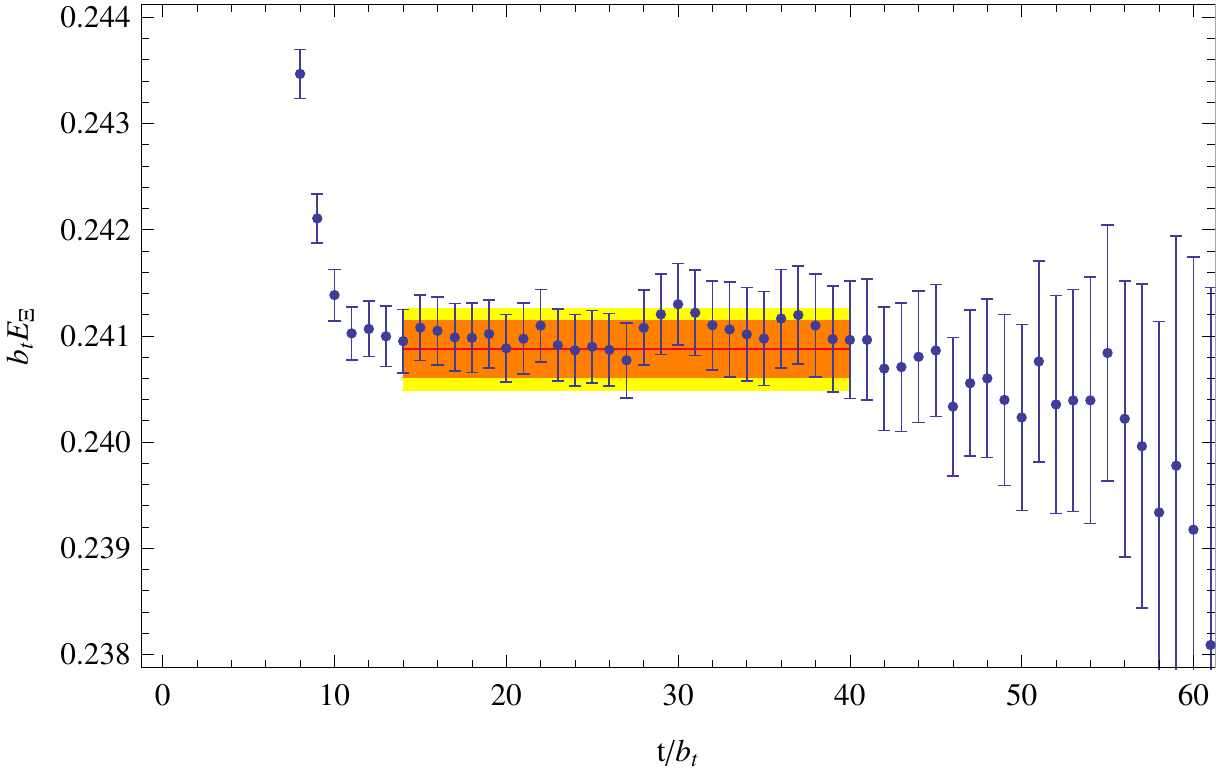}\ \ \ \ \ 
  \includegraphics[width=0.45\columnwidth]{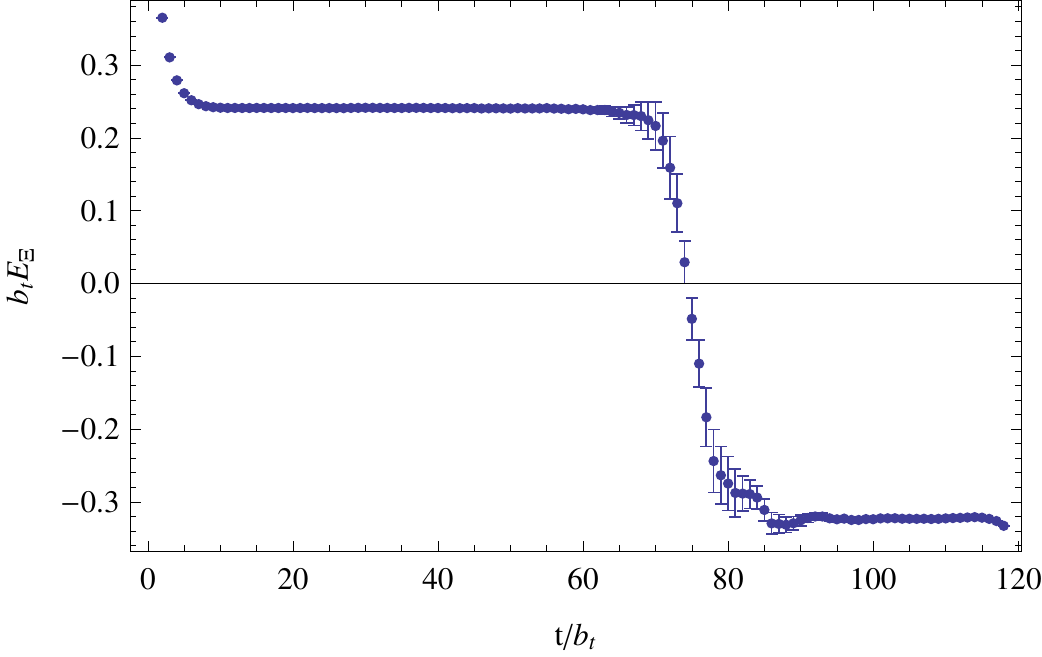}
  \caption{The EMP's associated with the diagonalized $\Xi^0$ correlation
    function, with $t_J=5$. The left panel, which also shows the fit to the plateau region,  
is a magnification  of the right panel.
}
  \label{fig:Xiemps}
\end{figure}
The effective mass plot (EMP) associated with the diagonalized nucleon
correlation function is shown in fig.~\ref{fig:Nemps}, and that
associated with the diagonalized $\Xi^0$ correlation function is shown
in fig.~\ref{fig:Xiemps}.  Extended and clean plateaus are observed
for both the nucleon and the $\Xi^0$, as discussed in detail in
Ref.~\cite{Beane:2009ky}.

\begin{figure}[!t]
  \centering
  \includegraphics[width=0.45\columnwidth]{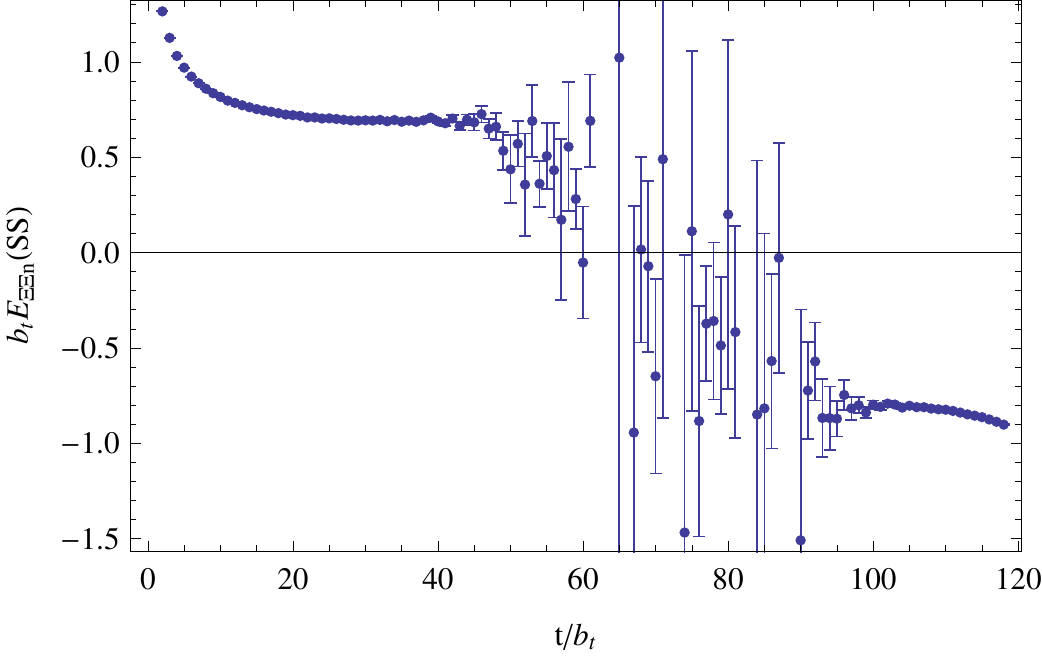}\ \ \ \ \ 
  \includegraphics[width=0.45\columnwidth]{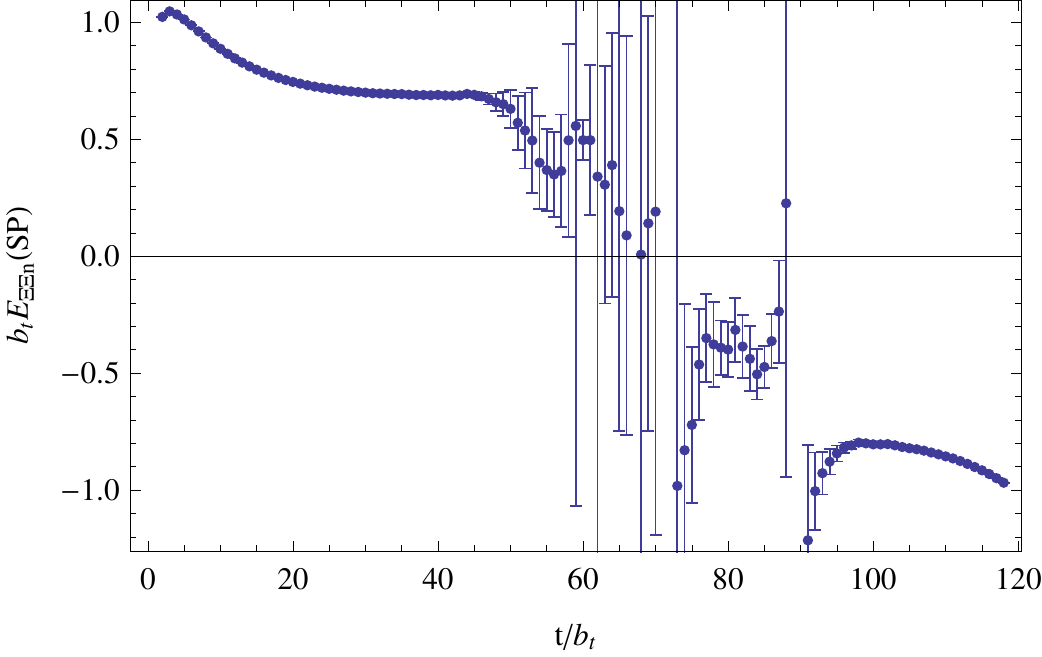}
  \includegraphics[width=0.45\columnwidth]{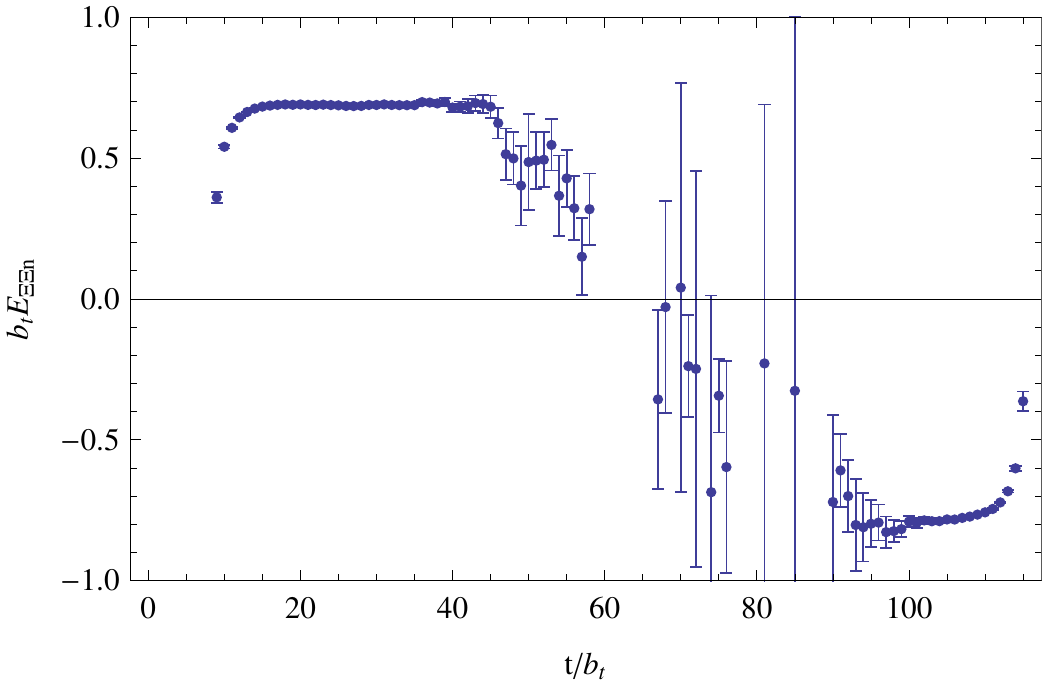}\ \ \ \ \ 
  \includegraphics[width=0.45\columnwidth]{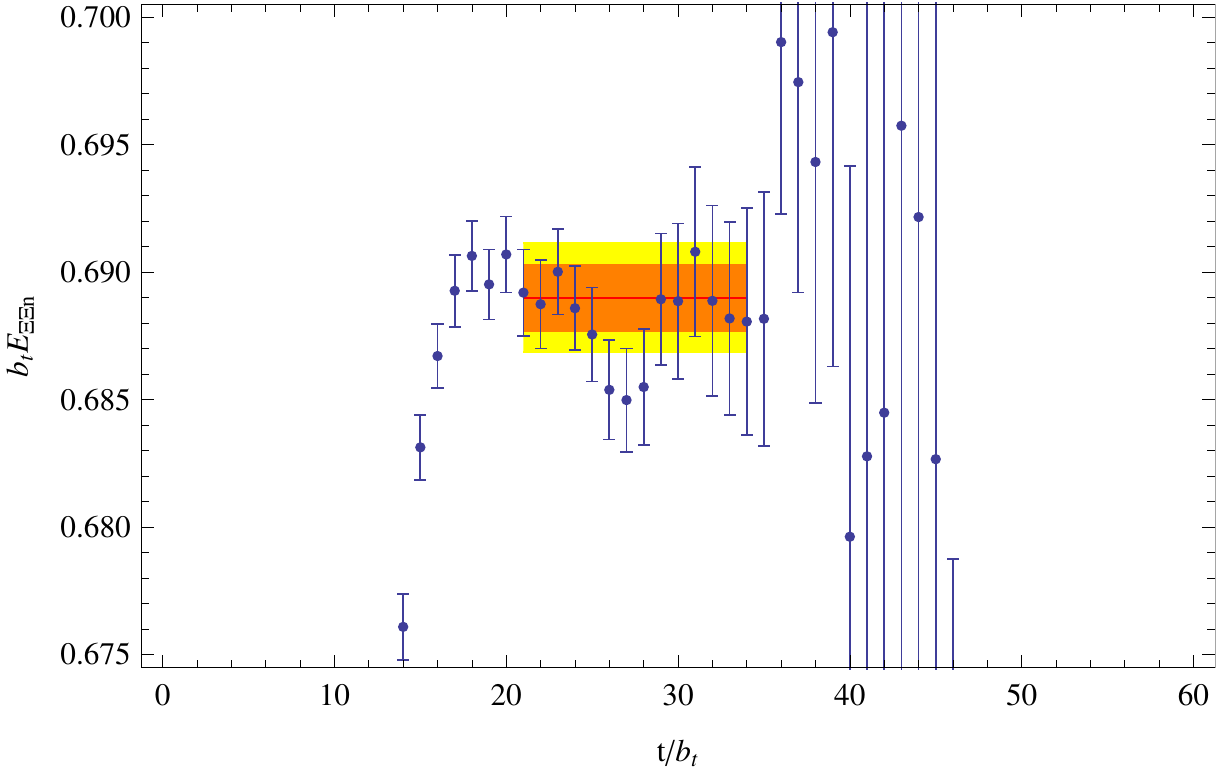}
  \caption{The EMP's associated with the SS, SP, and diagonalized
    $\Xi^0\Xi^0 n$ correlation function, with $t_J=5$. The lower right
    panel, which also shows the fit to the plateau region, is a
    magnification of the lower left panel.  }
  \label{fig:XiXinemps}
\end{figure}

The source and sink used to produce the $\Xi^0\Xi^0 n$ state is the
product of interpolating operators that have good overlap onto the
lowest-lying octet baryons, $\Xi^0$ and $n$.  However, since the
three-baryon eigenstates of the QCD Hamiltonian in the lattice volume,
or in nature, are not simple products of single baryon eigenstates,
this source and sink will couple (at some level) to all states with
the corresponding quantum numbers. This will be the case, no matter
how well the single hadron interpolating operators project to their
respective ground states.  We expect our correlation functions to have
significant contributions from nearby states, such as $\Xi^0\Xi^0 n$,
$\Xi^0\Xi^- p$, $\Xi^0\Lambda\Lambda$, $\Xi^0\Sigma^+\Sigma^-$,
$\Xi^0\Sigma^0\Sigma^0$, $\Xi^-\Sigma^+\Sigma^0$,
$\Xi^-\Sigma^+\Lambda$ with thresholds (neglecting interactions) of
$0.6893$, $0.6893$, $0.6858$, $0.7008$, $0.7008$, $0.7008$, $0.6933$
in lattice-units, respectively.  It is clear that the energy
eigenstates in the lattice volume will be mixtures of the different
states and from the above considerations, we expect to find four
relatively close energy-levels.~\footnote{It is interesting to note
  that non-strange channels will be simpler to analyze as the low
  energy spectrum will be less dense.}  In order to cleanly see these
nearby states, multiple correlation functions formed from different
sources and sinks, and more sophisticated analysis techniques will be
required.  As the goal of this work is not to provide detailed
spectroscopy of such states, but to demonstrate the feasibility of
studying such systems, we do not make efforts to identify the total
isospin of the ground state, and are content with identifying what
appears to be (with the current statistics) a single state. For
simplicity, we refer to this state as $\Xi^0\Xi^0n$.  The EMP's
associated with the SS, SP and diagonalized $\Xi^0\Xi^0 n$ correlation
functions are shown in fig.~\ref{fig:XiXinemps},
and show  a  plateau for a three-baryon state with energy
\begin{eqnarray}
E_{\Xi^0\Xi^0 n} & = & 0.6890(13)(17)(6)~{\rm l.u.}\ =\ 
3877.9\pm 6.9\pm 9.2\pm3.3~{\rm MeV}
\ \ ,\ \ \chi^2/{\rm d.o.f.} \ =\ 1.9
\ \ \ .
\label{eq:fitval3bdy}
\end{eqnarray}
The uncertainties in this result correspond to a statistical
uncertainty, a fitting uncertainty from the analysis presented
above, and an additional fitting uncertainty from comparison to
alternate analysis techniques using either multiple exponential fits
or other Prony methods (see Ref.\cite{Beane:2009ky} for details). An
additional 
uncertainty associated with the determination of  lattice scale
$b_s=0.1227(8)$ is not included.
Fig.~\ref{fig:XiXinemps} also shows a plateau for a  backward propagating negative
parity state, which is consistent with a four-body state with the
quantum numbers of $\Xi^0\Xi^0 n\pi$ (which ultimately will allow for the
calculation of pion interactions with multi-baryon systems).

\begin{figure}[!t]
  \centering
  \includegraphics[width=0.45\columnwidth]{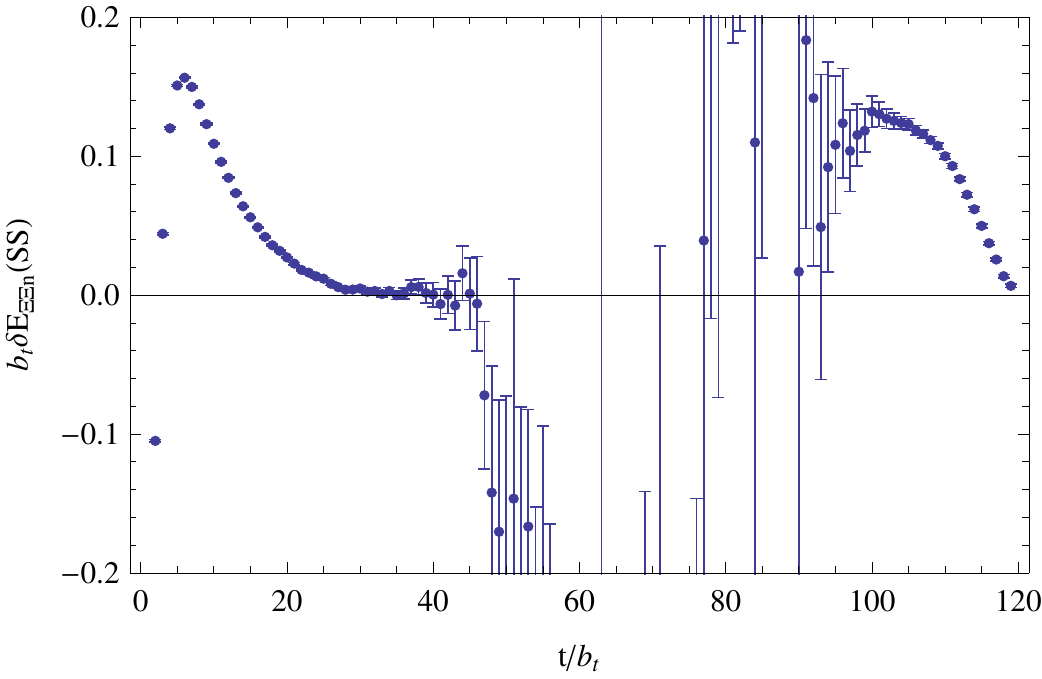}\ \ \ \ 
  \includegraphics[width=0.45\columnwidth]{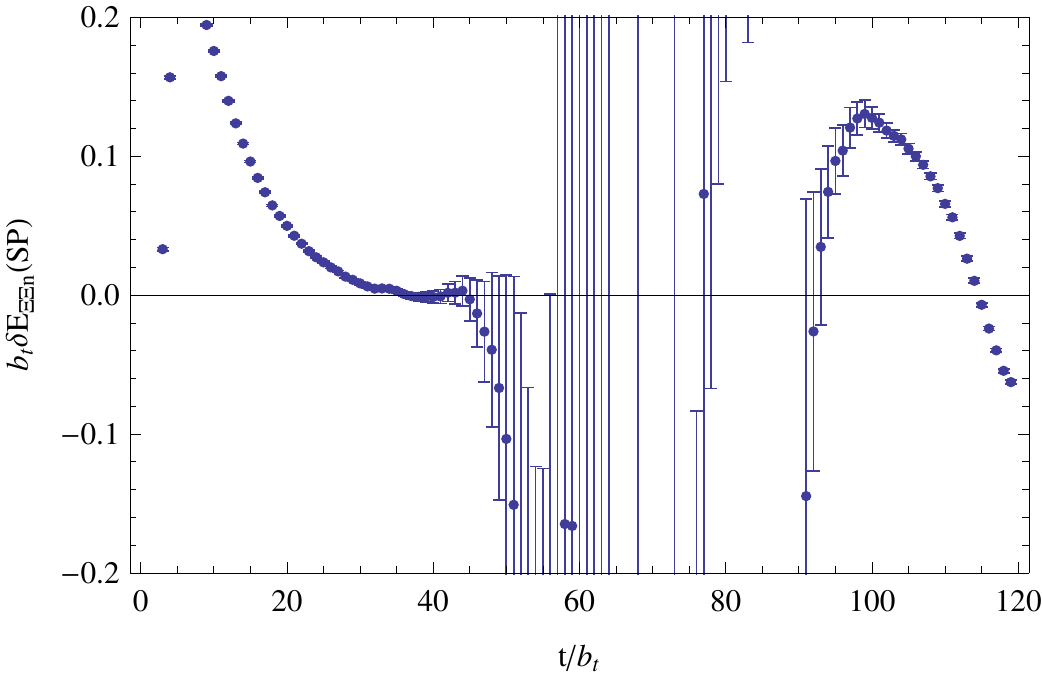} \\
  \includegraphics[width=0.45\columnwidth]{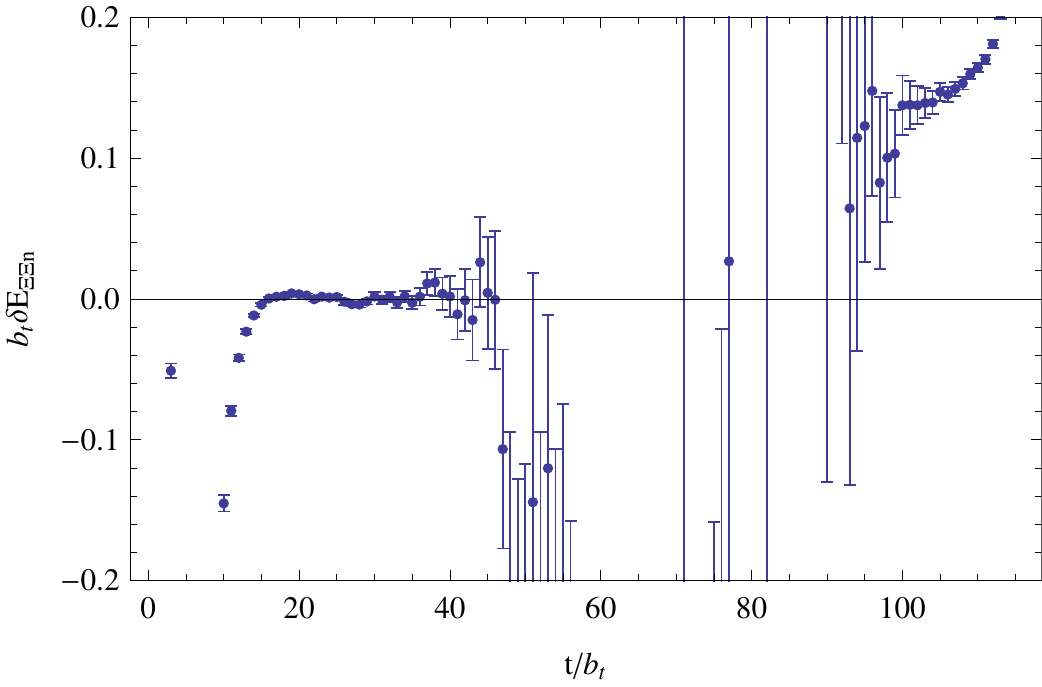}\ \ \ \ 
  \includegraphics[width=0.45\columnwidth]{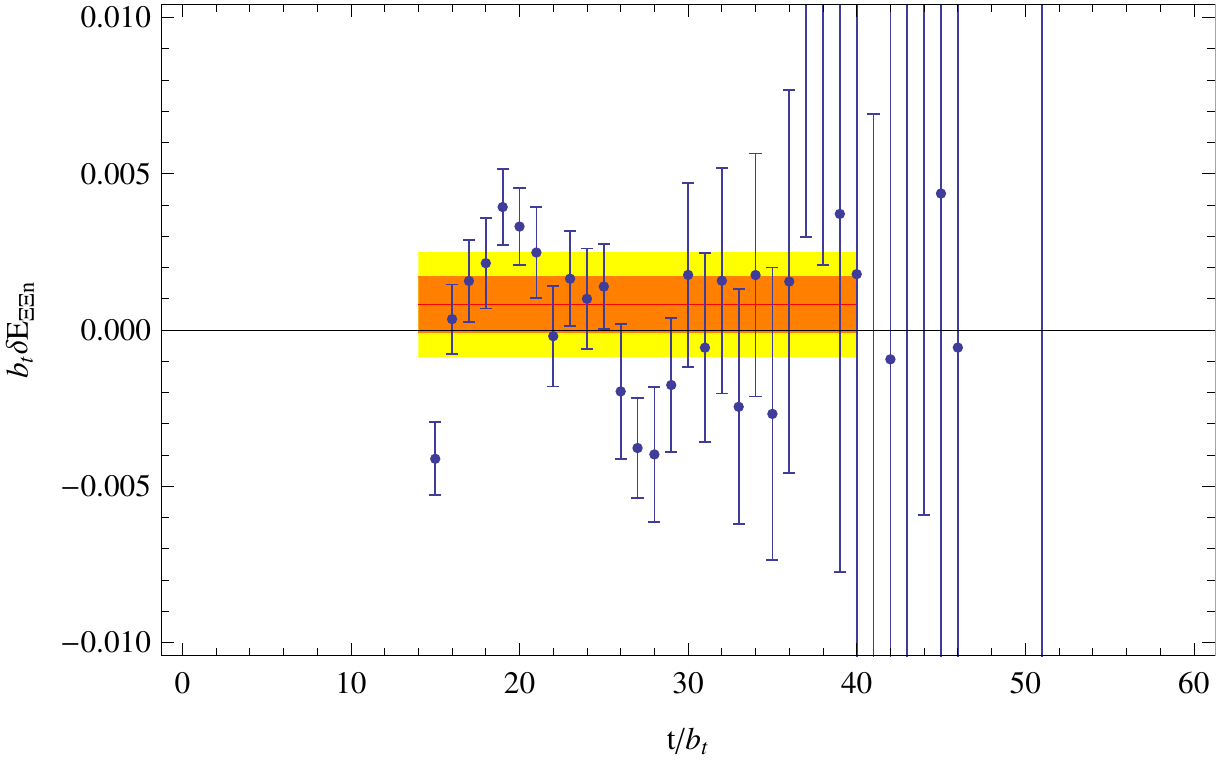}\\
  \caption{The EMP's for the energy-splitting 
associated with the ratio of SS, SP and diagonalized
    correlation functions, $\overline{G}_{\Xi^0\Xi^0 n}(t)$ as defined
    in eq.~(\protect\ref{eq:Gdef}) with $t_J=3$.  The lower right panel,
    which also shows the fit to the plateau region, is a magnification 
    of the lower left panel. 
 }
  \label{fig:XiXinRAT}
\end{figure}

As the diagonalized correlation functions are dominated by their
respective ground-states even at relatively short times, the
energy-splitting between the ground-state of the $\Xi^0\Xi^0 n$
system, and that of the two $\Xi^0$'s and a neutron can be found
efficiently by forming the ratio of the diagonalized correlation
functions, $\overline{C}_{\cal H}(t)$,
\begin{eqnarray}
\overline{G}_{\Xi^0\Xi^0 n}(t) & = & { \overline{C}_{\Xi^0\Xi^0 n}(t) \over 
    \overline{C}^2_{\Xi^0}(t)\  \overline{C}_{n}(t) }
\ \rightarrow\ 
{\cal A}_0\ 
e^{-\delta E_{\Xi^0\Xi^0 n} t}
 \ \ \ ,
  \label{eq:Gdef}
\end{eqnarray}
which at large times (for gauge-field configurations that are
infinitely long in the time-direction) tends to an exponential that
depends upon the energy-splitting $\delta E_{\Xi^0\Xi^0 n} =
E_{\Xi^0\Xi^0 n} - 2 M_{\Xi^0} - M_n$.  Fig.~\ref{fig:XiXinRAT} shows
the effective mass corresponding to $\overline{G}_{\Xi^0\Xi^0 n}(t)$,
along with the correlated fit to the plateau region.
The energy-splitting $\delta E_{\Xi^0\Xi^0 n}$ is determined to be
\begin{eqnarray}
\delta E_{\Xi^0\Xi^0 n} & = & 4.6\pm 5.0\pm 7.9\pm4.2~{\rm MeV}
\ \ ,\ \ \chi^2/{\rm d.o.f.} \ =\ 2.0
\ \ \ ,
\label{eq:fitval}
\end{eqnarray}
from fitting the time-interval from $t=21$ to $t=35$, using $t_J=3$,
which is consistent with zero (the splitting is computed relative to
the non interacting system of two $\Xi^0$s and a neutron for
convenience).  It is very encouraging that the uncertainty in the
energy-shift per baryon is $\sim 3~{\rm MeV}$, which is smaller than
the binding-energy per nucleon in typical nuclei, $B \sim 8~{\rm
  MeV}$, and not significantly larger than the binding-energy per
nucleon in the deuteron or triton at the physical values of the
light-quark masses.  The single energy-level fit to the EMP in
fig.~\ref{fig:XiXinRAT} has a $\chi^2/{\rm d.o.f.} \ =\ 2.0$,
indicating that there maybe additional structure in the correlation
function.  Including a second energy-level shifted by $\Delta E\sim
-0.004$ lattice units might provide a better description of the EMP,
and this would be consistent with the lower-energy state,
$\Xi^0\Lambda\Lambda$, that is expected to contribute to the four
low-lying eigenstates in the lattice-volume.  However, enhanced
statistics are required to determine if this is, in fact, the case.

At present, unlike the situation in multi-meson
systems~\cite{Beane:2007es,Detmold:2008fn,Detmold:2008yn}, the
analytical tools are not in place to use the above energy shift and
those of the associated two baryon systems to extract the parameters
describing the relevant two- and three-body interactions. While the
volume dependence of the simplest three-fermion systems has been
studied in Ref.~\cite{Luu:2008fg}, the mixing we expect between four
closely spaced states complicates the situation.

\section{Signal-To-Noise Ratios}
\label{sec:signaltonoise}

\noindent
The $\Xi^0\Xi^0 n$ calculation is  possible with our present
resources because there are  time-slices in the
correlation functions for which the signal-to-noise ratio is
approximately independent of time.  This is a region of time-slices
for which the correlation function that dictates the variance of the
signal is not yet dominated by its ``ground-state'', which for the
single-nucleon correlation function  is  three pions.  Given the importance
of this observation in Ref.~\cite{Beane:2009ky}, it is worth
re-stating and expanding upon it here.

As argued by Lepage~\cite{Lepage:1989hd}, correlation functions
involving one or more baryons exhibit statistical noise that increases
exponentially with Euclidean time.  In the case of a single positive
parity nucleon, the correlation function has the form
\begin{eqnarray}
  \langle \theta_{N}(t)\rangle &=&
  \sum_{\bf x}\ 
\Gamma_+^{\beta\alpha}\langle N^\alpha({\bf x},t) \overline{N}^{\beta} ({\bf 0},0)\rangle
  \ \rightarrow\ Z_N \ e^{-M_N t}
  \ \ ,
  \label{eq:Gfunproton}
\end{eqnarray}
where $N^\alpha({\bf x},t)$ is an interpolating field that has
non-vanishing overlap with the nucleon, $\Gamma_+$ is a positive
energy projector, and the angle brackets indicate
statistical averaging over measurements on an ensemble of
configurations.  
The variance of this correlation function is given by 
\begin{eqnarray}
  {\rm N}\  \sigma^2 & \sim & \langle \theta^{\dagger}_N(t)
  \theta_N(t)\rangle  - \langle \theta_{N}(t) \rangle^2 \nonumber \\
  & = & \sum_{\bf x,y} \Gamma_+^{\beta\alpha}\Gamma_+^{\gamma\delta}
  \langle N^\alpha({\bf x},t) \overline{N}^{\beta}({\bf y},t) N^\gamma({\bf 0},0)
  \overline{N}^{\delta}({\bf 0},0) \rangle\ \ -\ \langle \theta_{N}(t) \rangle^2
  \nonumber\\
  & \rightarrow & 
Z_{N\overline{N}} e^{-2 M_N t} - Z_N^2  e^{-2M_N t}
\ +\ 
Z_{3\pi}\  e^{-3 \mpi t}\ +\ ...
  \ \ \rightarrow \ Z_{3\pi}\  e^{-3 \mpi t}
  \ \ ,
  \label{eq:GGdaggerfunproton}
\end{eqnarray}
where all interaction energies have been neglected, and N is the number of
(independent) measurements (distinct from the nucleon field operator $N$).
Therefore, at large times,  the
noise-to-signal ratio behaves as
\begin{eqnarray} {\sigma\over\overline{x}} & = & {\sigma (t)\over
    \langle \theta(t) \rangle } \sim {1\over \sqrt{\rm N}} \ e^{\left( M_N
      - {3\over 2} \mpi\right) t} \ \ .
  \label{eq:NtoSproton}
\end{eqnarray}
More generally, for a system of $A$ nucleons, the noise-to-signal
ratio behaves as
\begin{eqnarray} {\sigma\over\overline{x}} & & \sim {1\over \sqrt{\rm N}}
  \ e^{A \left( M_N - {3\over 2} \mpi\right) t} \ \ 
  \label{eq:NtoSnucleus}
\end{eqnarray}
at large times.  The degradation of the signal-to-noise ratio on
gauge-field configurations of finite temporal extent is exponentially
more rapid than that given in Eq.~(\ref{eq:NtoSnucleus}) due to the
presence of thermal states, as discussed in Ref.~\cite{Beane:2009ky}.

From the signals and variances that we have measured in the one-, two-
and three-baryon sectors, it is clear that there is a suppression of
the overlap onto the three-meson state from the $N\overline{N}$ source
and sink (the variance correlation function of
Eq.~(\ref{eq:GGdaggerfunproton})), as encapsulated in the factor
$Z_{3\pi}$.  If $Z_{3\pi} \ll Z_{N\overline{N}},\, Z_N^2$ there will
be a number of time-slices, near the source of the correlation
function, for which the noise-to-signal ratio behaves as
\begin{eqnarray}
 {\sigma\over\overline{x}} & & \sim {1\over \sqrt{\rm N}}
 \ \ ,
  \label{eq:NtoSnucleusGOLDEN}
\end{eqnarray}
and does not depend exponentially upon time, or the differences of
hadron masses.  The correlation functions we have constructed lead to
an implicit suppression of $Z_{3\pi}$ compared to $Z_{N\overline{N}}$
and $Z_N^2$, due to the fact that the overlap onto the three-meson
state, or any meson state, is strongly suppressed when the sinks
$N^\alpha({\bf x},t)$ and $\overline{N}^{\beta}({\bf y},t)$ do not
overlap within a volume approximately defined by the pion Compton
wavelength.  Therefore, summing independently over the volumes for
$N^\alpha({\bf x},t)$ and $\overline{N}^{\beta}({\bf y},t)$ leads to a
suppression factor that scales with the spatial lattice
volume~\footnote{This has been noted independently by D.~B.~Kaplan
  \cite{DBK_PC}.} as $Z_{3\pi}/Z_{N\overline{N}} \sim 1/\Lambda^3 L^3
$.  The pion mass dictates the weakest suppression, and hence we set
$\Lambda=\mpi$ for the estimates that follow (if the width, $w$, of
the smearing of the source and sink is larger than the pion Compton
wavelength, then $\Lambda\sim 1/w$). Results consistent with this
volume scaling have been found explicitly in calculations of single
baryon energies using domain-wall fermions on MILC gauge
configurations.

By generalizing this argument to systems composed of $A$
nucleons\footnote{This simple argument holds for $A\leq 4$ nucleons and
  generalizes simply to $A\leq16$ octet baryons. 
Further generalizations of this argument accounting for
  Fermi statistics are straightforward.}  where each
interpolating field is projected to zero momentum, the noise correlation
function is expected to behave parametrically as
\begin{eqnarray}
  \label{eq:Anoise}
  {\rm N}\ \sigma^2 &\sim&
 \frac{(A!)^2}{(\mpi L)^{3A}} Z_A \ e^{-3A\, \mpi t}
+\frac{((A-1)!)^2}{(\mpi L)^{3(A-1)}} Z_{A-1} \ e^{-(2M_N+3(A-1) \mpi)t}
+\ldots \nonumber 
\\
&&
+\frac{A^2}{(\mpi L)^3}Z_1 \ e^{-(2(A-1)M_N+3 \mpi) t}
\ +\  Z_0 \ e^{-2A\, M_N t} +\ldots\,,
\end{eqnarray}
where we have made explicit the parametric dependence of the overlap
factors on the baryon number and spatial volume. The dependence on $A$
arises from the number of ways that $N$ and $\overline{N}$ sink
operators can overlap to form one or more three-pion contributions to
the correlation function.  Provided that the spatial volume is large
compared to the Compton wavelength of the pion, $\mpi L\gg1$, there
will be a range of time-slices in which only the last two terms in
Eq.~(\ref{eq:Anoise}) are important. In this region, the
signal-to-noise ratio in the multi-baryon correlation function does
not degrade exponentially faster that the signal-to-noise ratio in the
single baryon correlation function.  Instead,
\begin{eqnarray}
  \label{eq:1}
  \frac{\sigma}{\overline{x}} &\sim& 
{1\over\sqrt{\rm N}}
\left(\ Z\ +\ Z^\prime\ {A^2\over m_\pi^3 L^3}\
  e^{(M_N-\frac{3}{2}\mpi)t}
\ +\ ...
\right)
\ \ \ ,
\end{eqnarray}
where $Z$ and $Z^\prime$ are ${\cal O}(1)$ ratios of overlap
factors. Consequently, the signal-to-noise ratio starts degrading
exponentially only after time-slice $t_{\rm noise}$, which has
parametric dependence
\begin{equation}
  \label{eq:3}
  t_{\rm noise} \sim \frac{2}{2M_N-3\mpi}\ \ln\left[\frac{ \mpi^3 L^3}{A^2}\right]
\ \ \ .
\end{equation}
It is important to note that  $t_{\rm noise}$
depends only logarithmically on the number of baryons, and hence it is
conceivable that plateaus may be found in the EMP's of systems containing four
or more baryons with the current number of measurements if the contractions
are performed.

%
%
\begin{figure}[!t]
\center
    \includegraphics[width=0.9\textwidth]{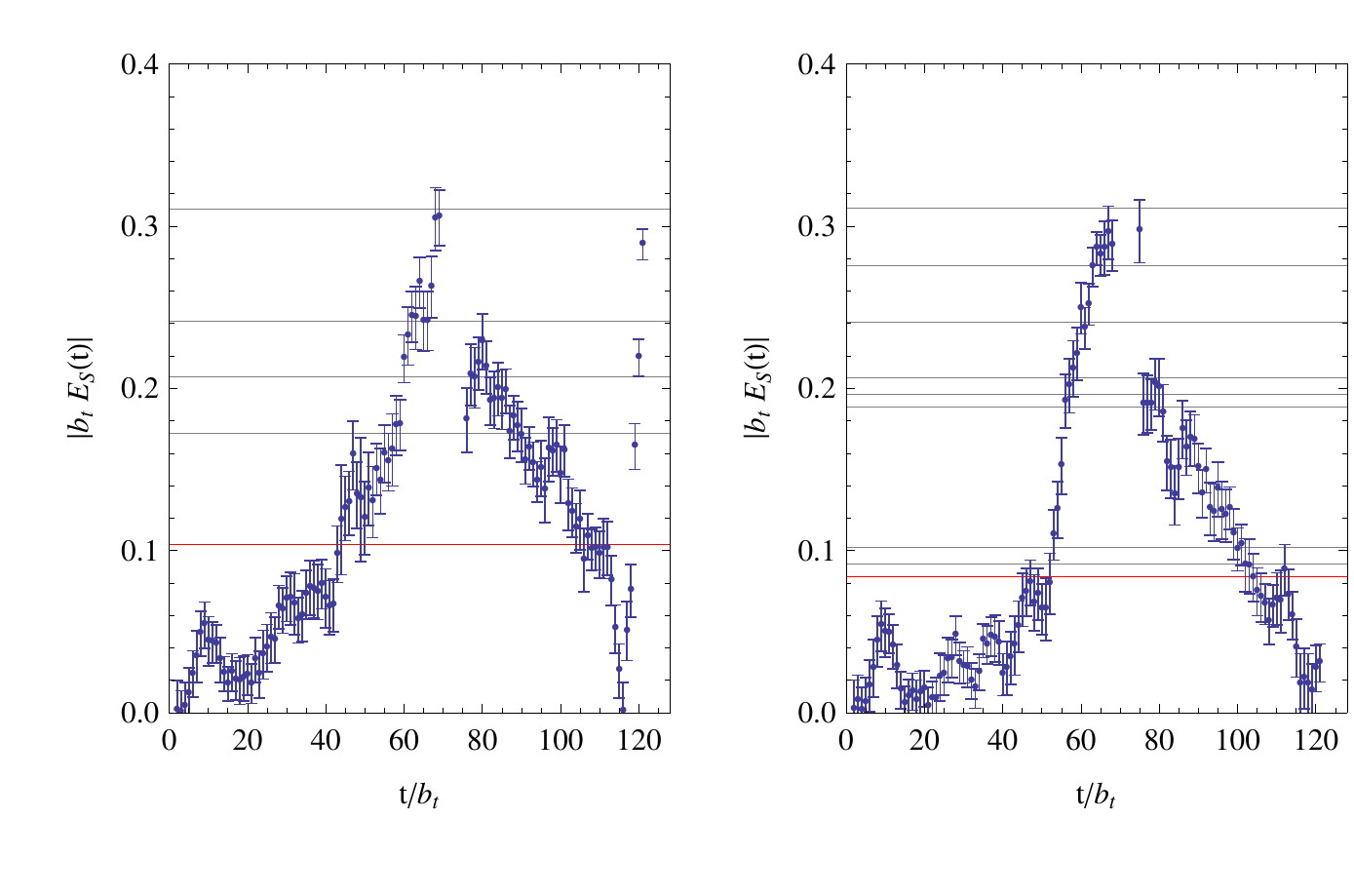}
  \caption{\label{fig:StoN}  
The energy-scales associated with the signal-to-noise ratios for the nucleon
(left panel) and $\Xi$ (right panel), as defined in
eq.~(\protect\ref{eq:Estondefn}).
The horizontal lines in the left panel correspond to 
the energy scales $m_p-{3\over 2} \mpi$, $m_p-{1\over 2} \mpi$,
    $m_p$, $m_p+{1\over 2} \mpi$, and $m_p+{3\over 2} \mpi$ (from
    lowest energy to highest energy).
The horizontal lines in the right panel correspond to 
the energy scales $m_\Xi-{3\over 2} m_\eta$, $m_\Xi-\mK-{1\over
      2} m_\eta$, $m_\Xi-m_\eta - {1\over 2} \mpi$, $m_\Xi-{1\over 2}
    m_\eta$, $m_\Xi-\mK + {1\over 2} m_\eta$, $m_\Xi-{1\over 2} \mpi$,
    $m_\Xi$, $m_\Xi+{1\over 2} \mpi$, and $m_\Xi+m_\eta -{1\over 2}
    \mpi$ (from lowest energy to highest energy). 
}
\end{figure}
In order to investigate the signal-to-noise ratio in the correlation functions
of interest, it is useful to form the
effective noise-to-signal plot~\cite{Beane:2009ky}, in analogy with
the EMPs.  On each time slice, the quantity
\begin{eqnarray} {\cal S}(t) & = & {\sigma (t)\over\overline{x}(t)} \
  \ \ ,
  \label{eq:stondefn}
\end{eqnarray}
is formed, from which the energy governing the exponential behavior
can be extracted via
\begin{eqnarray}
  E_{\cal S}(t;t_J) & = & 
  {1\over t_J}\ \log\left({ {\cal S}(t+t_J) \over {\cal S}(t)}\right)
  \ \ \ .
  \label{eq:Estondefn}
\end{eqnarray}
If the correlation function is dominated by a single state, and a
single energy-scale determines the behavior of the noise-to-signal
ratio, the quantity $E_{\cal S}(t;t_J)$ will be independent of both
$t$ and $t_J$.

In fig.~\ref{fig:StoN}, the energy scales of the noise-to-signal ratio
are shown for the nucleon and $\Xi^0$. As discussed previously in
Ref.~\cite{Beane:2009ky}, it is clear that for $t\agt45$ this scale is
significantly greater than the asymptotic estimate of Lepage (the
lowest horizontal line in each figure) because of thermal states
involving propagation around the temporal extent of the lattice.  It
is also clear that even the simple Lepage scaling does not set in for
many time-slices corresponding to a large window in which the signal
is statistically clean.  Figure~\ref{fig:StoN} indicates that the
suppression of mesonic intermediate states is stronger for the $\Xi$
correlation function than for the nucleon correlation function, as
evidenced by the energy-scale of the signal-to-noise ratio remaining
small for longer times.

\begin{figure}[!t]
   \center
    \includegraphics[width=0.9\textwidth]{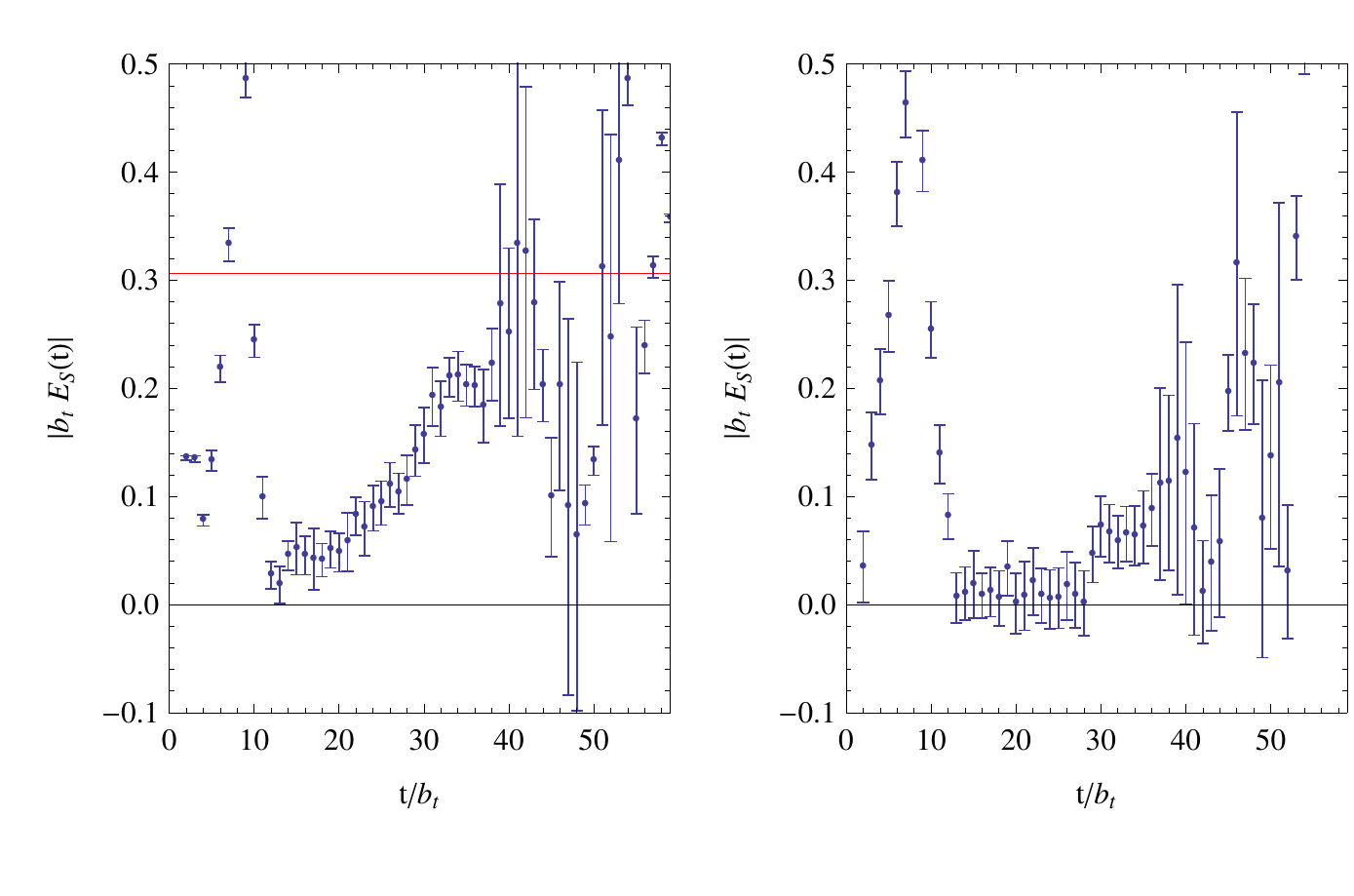}\
  \caption{\label{fig:StoN3} The left panel shows the energy-scales
    associated with the signal-to-noise ratios for the $\Xi^0\Xi^0 n$
    correlation function, as defined in
    eq.~(\protect\ref{eq:Estondefn}).  The horizontal line corresponds
    to $m_N + 2 m_\Xi - 2 m_\eta - {5\over 2} m_\pi $, the asymptotic
    energy-scale in a lattice with infinite temporal extent.  The
    right panel shows the difference between the signal-to-noise
    energy scales of the diagonalized $\Xi^0\Xi^0 n$ correlation
    function and that of the nucleon and twice that of the $\Xi$
    correlation function.  }
\end{figure}

The energy-scale associated with the noise-to-signal ratio of the
diagonalized $\Xi^0\Xi^0 n$ correlation function is shown in the left
panel of fig.~\ref{fig:StoN3}.
%
%
While degrading exponentially, the signal-to-noise ratio of the
three-body correlation function, is exponentially better in the
plateau region $t \lsim 32$, than expected based upon the
arguments of Lepage, consistent with the expectations based upon the
behavior of the signal-to-noise ratio of the nucleon and $\Xi$. The
right panel of fig.~\ref{fig:StoN3} shows that the energy-scale
associated with the signal-to-noise ratio in the $\Xi^0\Xi^0 n$
correlation function is consistent with the simple sum of the
energy-scales from the single baryon correlation functions (within
statistical uncertainties of the calculation).

In creating sources and sinks for correlation functions, a great deal
of attention is paid to optimizing the overlap onto the states of
interest.  Variational techniques
\cite{Michael:1985ne,Luscher:1990ck}, the matrix-Prony method, and
related approaches make use of sources and sinks with substantial, but
different, overlaps onto the states of interest to enable a
diagonalization to the eigenstates in the lattice-volume (up to
exponentially suppressed contributions).  For multi-baryon systems,
the results of this work make clear that an equally important
component of source and sink optimization is to minimize the overlap
onto the lightest states that contribute to the variance of the
correlation function.  This will also be true for the extraction of
the properties of excited single particle states.

\begin{figure}[!t]
  \centering
  \includegraphics[width=0.7\columnwidth]{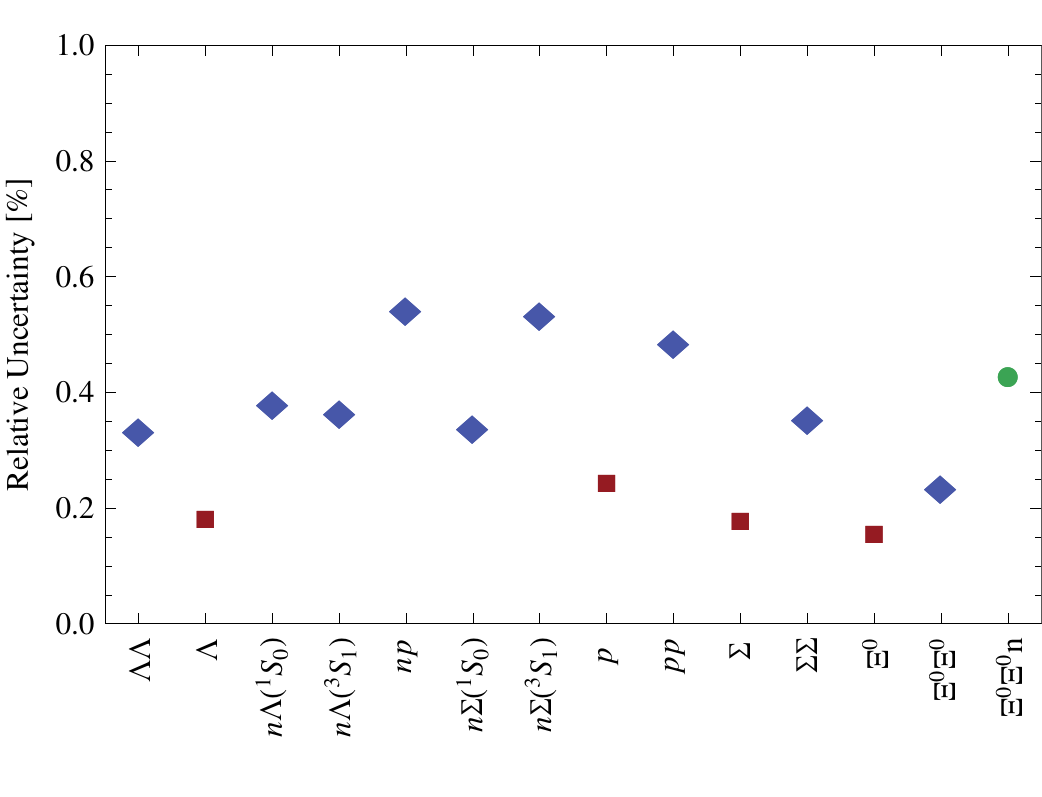} \\
  \includegraphics[width=0.7\columnwidth]{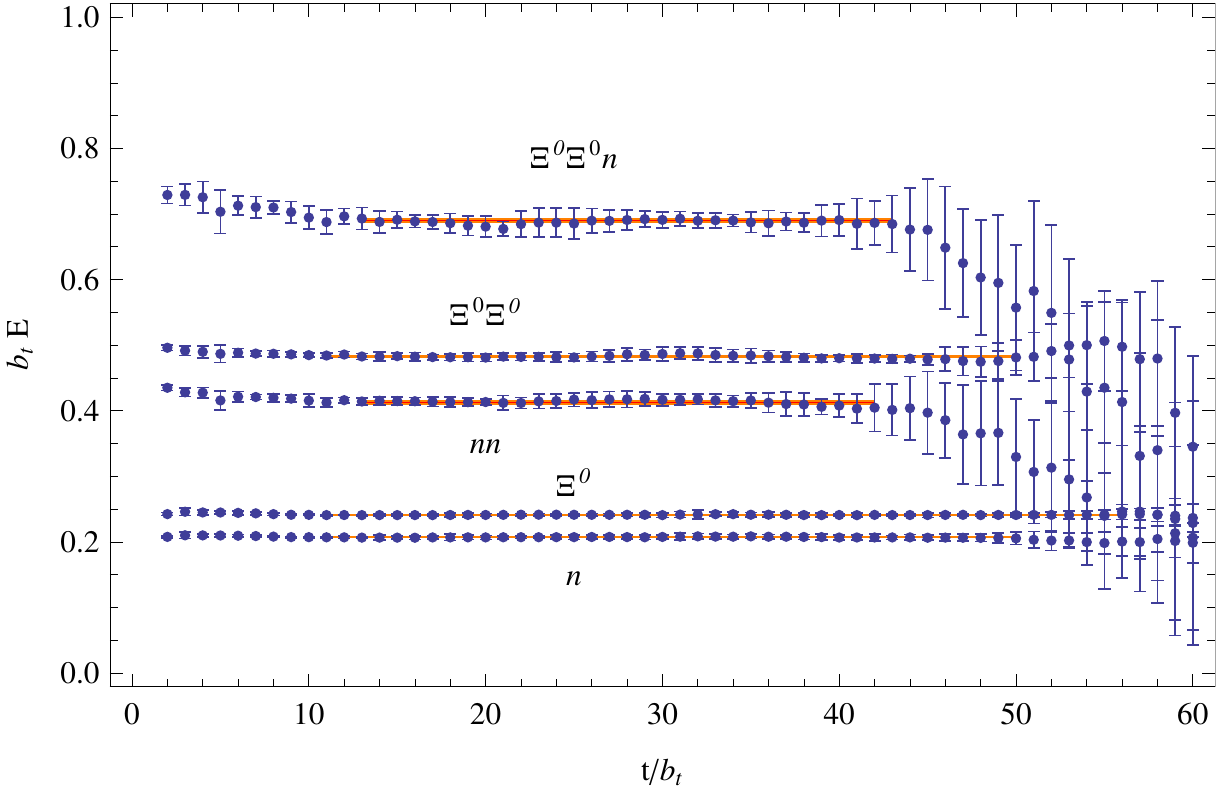}
  \caption{The upper panel shows the relative uncertainties in the
    extraction of one-, two- and three-baryon ground state energies. The
  lower panel shows the corresponding EMPs for some of these systems obtained
  with the Matrix-Prony method.}
  \label{fig:relunc}
\end{figure}

Figure \ref{fig:relunc} shows a comparison of the relative
uncertainties (statistical and systematic uncertainties are added in
quadrature and normalized by the mean value of the measurement) in the
extraction of the ground state hadron energy for a selection of one-,
two- and three-baryon systems using the measurements for which
$\Xi^0\Xi^0n$ contractions exist. We find that this quantity is
approximately constant, due to there being a sufficiently large window
of time-slices for which the signal-to-noise ratio does not degrade
exponentially. The extent of this window is empirically seen to
decrease with the number of baryons as shown in the lower panel of
fig~\ref{fig:relunc} where the matrix-Prony effective
energies\footnote{These EMPS and the central values of these
  extractions slightly differ from the extractions presented in
  figs.~\ref{fig:Nemps}, \ref{fig:Xiemps}, and \ref{fig:XiXinemps},
  but are consistent.} of exemplary one-, two- and three-baryon
systems are shown (detailed analysis of the two-baryon sector will
appear in future work \cite{twobody}). This result is consistent with
the scaling anticipated in Eq.~(\ref{eq:3}).

\section{The Triton Channel}
\noindent
One of the main motivations for the present work is to show that
three-baryon (and beyond) calculations can be done at unphysical quark
masses with present day resources.  We have shown the results for
channels that couple to $\Xi^0\Xi^0 n$ sources and sinks as their
correlation functions are some of the least computationally expensive,
requiring the calculation of only 288 Wick contractions.  A
significantly more computationally expensive, but physically more
interesting, channel is that of the triton ($pnn$) for which there are
2880 Wick contractions.  We do not have the computational resources
available to perform the $pnn$ contractions on all of the 260,000
measurements we have made, and to date have only performed $\sim
9,200$ measurements of the smeared-smeared correlation function.
%
%
\begin{figure}[!t]
 \center
    \includegraphics[width=0.7\textwidth]{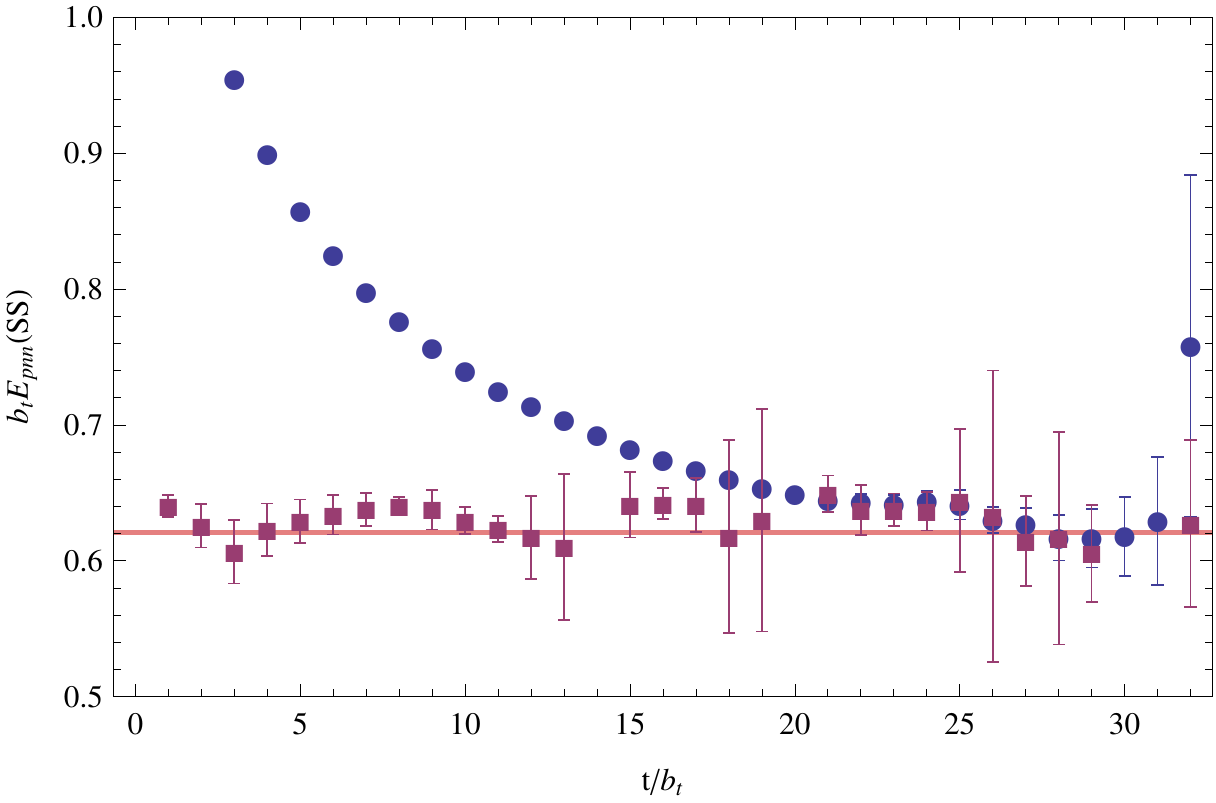}
    \caption{\label{fig:triton} The standard EMP for the
      smeared-smeared $pnn$ correlation function is shown (circles)
      along with the generalized EMP resulting from a 3-exponential
      Prony analysis~\protect\cite{Beane:2009ky} of the data
      (squares).  The horizontal light band corresponds to three-times
      the nucleon mass.  }
\end{figure}
The generalized EMP resulting from a 3-exponential Prony analysis is
shown in fig.~\ref{fig:triton}, and given the relatively large
uncertainties, we do not present a value for the ground-state energy.
At the physical pion mass one expects to find a negatively shifted
state corresponding to the triton. To conclude that such a signal
corresponded to a bound state would require further studies showing
exponentially suppressed sensitivity to the lattice volume in contrast
to continuum states.

\section{Conclusions \label{sec:conclusions}}
\noindent
In this work, we have presented the first Lattice QCD calculations of
a three-baryon state, focusing on a system with the quantum numbers of
$\Xi^0\Xi^0n$.  We find a ground state energy of $E_{\Xi^0\Xi^0
  n}=3877.9\pm 6.9\pm 9.2\pm3.3$~MeV corresponding to an energy shift from
the free three-baryon system of $\delta E_{\Xi^0\Xi^0 n}= 4.6\pm
5.0\pm 7.9\pm4.2~{\rm MeV}$.  Our high-statistics analysis of the behavior
of the signal-to-noise ratio of single- and multiple-baryon
correlation functions indicates that there is a window of time-slices
for which the signal-to-noise ratio does not degrade exponentially.
This implies that multi-baryon correlation functions can be calculated
in this time-interval with significantly less computational resources
than previously estimated, and we demonstrate that this is indeed the
case in the three-baryon sector.  The signal-to-noise ratio does not
depend exponentially upon the number of baryons in this time-interval,
however, the length of this window decreases logarithmically with the
number of baryons.

We have focused only on the state(s) that couples to the $\Xi^0\Xi^0
n$ interpolating-operator simply due to limited computational
resources.  In the past it has been the case that gauge-field
generation has required the majority of LQCD resources, but this is no
longer true for precise calculations of baryonic observables.  The
resources required to perform the large number of measurements
required for nuclear systems is significantly greater than that
required for gauge-field generation.  This situation will improve as
more effort is put into algorithmic improvements for contractions, in
the same way that the use of deflation~\cite{Stathopoulos:2007zi} and
other techniques have greatly reduced the resources required for
propagator generation. Work in this direction is in progress. Given the observed behavior of the signal-to-noise ratio,
we hope to be able to identify at least the ground state in 
systems of four and five baryons.

As the central goal for applications of lattice QCD to nuclear physics
is the calculation of nuclei and their interactions, we have also
presented the first calculations of the correlation function that
would contain the triton if the calculations were at the physical pion
mass.  The statistics are very limited compared with the $\Xi^0\Xi^0
n$ correlation function, but it is encouraging to see that there is a
clear plateau visible in the effective mass (within somewhat large
uncertainties).

The increase by more than one order of magnitude in the number of
measurements performed on a given ensemble of gauge-field
configurations has given rise to a new understanding of how to pursue
nuclear physics processes with Lattice QCD.  Source and sink
optimization involves two considerations (maximal overlap onto the
baryon states and minimum overlap onto the mesonic states in the
correlation function dictating the variance of the baryon correlation
functions) to make optimal use of available resources.  It is clear
that, at unphysical values of the quark masses, high statistics
calculations can be used to explore multi-nucleon systems (perhaps
beyond $A=5$) with present day resources.

\section{Acknowledgments}

\noindent 
We thank R.~Edwards and B.~Joo for help with the QDP++/Chroma
programming environment~\cite{Edwards:2004sx} and D. B. Kaplan for
discussions.  KO would like to thank A. Stathopoulos useful
discussions on numerical linear algebra issues and for his
contribution in the development of the EigCG algorithm
\cite{Stathopoulos:2007zi}. EigCG development was supported in part by
NSF grant CCF-0728915. We also thank the Hadron Spectrum Collaboration
for permitting us to use the anisotropic gauge-field configurations,
and extending the particular ensemble used herein.  We gratefully
acknowledge the computational time provided by NERSC (Office of
Science of the U.S. Department of Energy, No. DE-AC02-05CH11231), the
Institute for Nuclear Theory, Centro Nacional de Supercomputaci\'on
(Barcelona, Spain), Lawrence Livermore National Laboratory, and the
National Science Foundation through Teragrid resources provided by the
National Center for Supercomputing Applications and the Texas Advanced
Computing Center.  Computational support at Thomas Jefferson National
Accelerator Facility and Fermi National Accelerator Laboratory was
provided by the USQCD collaboration under {\it The Secret Life of a
  Quark}, a U.S. Department of Energy SciDAC project ({\tt
  http://www.scidac.gov/physics/quarks.html}).  The work of MJS 
was supported in part by the U.S.~Dept.~of Energy under Grant
No.~DE-FG03-97ER4014.  The work of KO and WD was supported in part by
the U.S.~Dept.~of Energy contract No.~DE-AC05-06OR23177 (JSA) and DOE
grant DE-FG02-04ER41302. KO and AWL were supported in part by the
Jeffress Memorial Trust, grant J-813 and  DOE OJI grant
DE-FG02-07ER41527. The work of SRB and AT was supported in part by the
National Science Foundation CAREER grant No.  PHY-0645570.  Part of
this work was performed under the auspices of the US DOE by the
University of California, Lawrence Livermore National Laboratory under
Contract No. W-7405-Eng-48.  The work of AP is partly supported by the
Spanish Consolider-Ingenio 2010 Programme CPAN CSD2007-00042, by
grants Nos. FIS2008-01661 from MICINN (Spain) and FEDER and
2005SGR-00343 from Generalitat de Catalunya, and by the EU contract
FLAVIAnet MRTN-CT-2006-035482.

%
%

\end{document}